\documentclass[12pt]{article}
\usepackage{pdproc}

  \makeatletter 
  \def\@cite#1{[#1]} 
  \makeatother    
\usepackage{amssymb,cite}
\usepackage{epsf}
\usepackage{epsfig}

\newcommand{\lsim}{
\mathrel{\hbox{\rlap{\hbox{\lower4pt\hbox{$\sim$}}}\hbox{$<$}}}}
\newcommand{\gsim}{
\mathrel{\hbox{\rlap{\hbox{\lower4pt\hbox{$\sim$}}}\hbox{$>$}}}}

\def\epe{\varepsilon'/\varepsilon}

\newcommand{\be}{\begin{equation}}
\newcommand{\ee}{\end{equation}}
\newcommand{\bi}{\begin{itemize}}
\newcommand{\ei}{\end{itemize}}

\def\kpn{K^+\rightarrow\pi^+\nu\bar\nu}

\def\klpn{K_{\rm L}\rightarrow\pi^0\nu\bar\nu}

\begin{document}

%%%%%%%%%%%%%%%%%%%%%

\begin{titlepage}
\vspace*{-0.5truecm}

\begin{flushright}
CERN-PH-TH/2004-237\\
TUM-HEP-569/04\\
MPP-2004-154\\
hep-ph/0411373
\end{flushright}

\vspace*{0.3truecm}

\begin{center}
\boldmath
{\Large{\bf $B\to\pi\pi$, New Physics in
$B\to\pi K$ and \\[3mm] Rare $K$ and $B$ Decays$^\dagger$}}

\vspace{0.3truecm}

\unboldmath
\end{center}

\vspace{0.5truecm}

\begin{center}
{\bf Andrzej J. Buras,${}^a$ Robert Fleischer,${}^b$ 
Stefan Recksiegel${}^a$ and Felix Schwab${}^{c,a}$}
 
\vspace{0.4truecm}

${}^a$ {\sl Physik Department, Technische Universit\"at M\"unchen,
D-85748 Garching, Germany}

\vspace{0.2truecm}

${}^b$ {\sl Theory Division, Department of Physics, CERN, 
CH-1211 Geneva 23, Switzerland}

\vspace{0.2truecm}

 ${}^c$ {\sl Max-Planck-Institut f{\"u}r Physik -- Werner-Heisenberg-Institut,
 D-80805 Munich, Germany}

\end{center}

\vspace{0.7cm}
%\begin{abstract}
\vspace{0.2cm}\noindent
We summarize a recent strategy for a global analysis of the $B\to\pi\pi, 
\pi K$ systems and rare decays. We find that the present $B\to \pi\pi$ 
and $B\to \pi K$ data cannot be simultaneously described in the 
Standard Model. In a simple extension in which new physics 
enters dominantly through $Z^0$ penguins with a CP-violating phase, 
only certain $B\to \pi K$ modes are affected by new physics. The 
$B\to \pi \pi$ data can then be described entirely within the Standard Model 
but with values of hadronic parameters that reflect large non-factorizable 
contributions. Using the $SU(3)$ flavour symmetry and plausible dynamical 
assumptions, we can then use the $B\to\pi\pi$ decays to fix the hadronic 
part of the $B \to \pi K$ system and make predictions for various observables 
in the $B_d \to \pi^{\mp} K^{\pm}$ and $B^{\pm} \to \pi^{\pm} K$ decays that 
are practically unaffected by electroweak penguins. The data on the  
$B^{\pm} \to \pi^0 K^{\pm}$ and $B_d \to \pi^0 K$ modes allow us then 
to determine the electroweak penguin component which differs from the 
Standard Model one, in particular through a large additional CP-violating 
phase. The implications for rare $K$ and $B$ decays are spectacular. In 
particular, the rate for $\klpn$ is enhanced by one order of magnitude, the 
branching ratios for $B_{d,s}\to \mu^+ \mu^-$ by a factor of five, and 
$\mathrm{BR}(K_{\rm L}\to \pi^0 e^+e^-, \pi^0 \mu^+ \mu^-)$
%$\mathrm{BR}(K_{\rm L}\to \pi^0 \mu^+ \mu^-)$ 
by factors of three. 
%We give a list of predictions for CP asymmetries in $B\to \pi \pi$ and 
%$B\to \pi K$ systems and for the rates of rare decays. 
%The overall agreement with the data is very good.
%\end{abstract}

\vspace*{0.5truecm}
\vfill
\noindent
$^\dagger$ Talk presented by A.~J.~Buras at SUSY 04, June 17-23, 2004,
Tsukuba, Japan\\[3mm]
November 2004

\end{titlepage}

\mbox{}
\pagebreak
\setcounter{page}{0}

%%%%%%%%%%%%%%%%%%%%%%%%%%

\renewcommand{\thefootnote}{\alph{footnote}}

\title{\boldmath $B\to\pi\pi$, New Physics in
$B\to\pi K$ and Rare $K$ and $B$ Decays}

\author{Andrzej J. Buras${}^a$, Robert Fleischer${}^b$, 
Stefan Recksiegel${}^a$ and Felix Schwab${}^{c,a}$}
 
\address{${}^a$ {Physik Department, Technische Universit\"at M\"unchen,
D-85748 Garching, Germany}\\
${}^b$ {Theory Division, Department of Physics, CERN, 
CH-1211 Geneva 23, Switzerland}\\
${}^c$ {Max-Planck-Institut f{\"u}r Physik -- Werner-Heisenberg-Institut,
 D-80805 Munich, Germany}}

\abstract{
We summarize a recent strategy for a global analysis of the $B\to\pi\pi, 
\pi K$ systems and rare decays. We find that the present $B\to \pi\pi$ 
and $B\to \pi K$ data cannot be simultaneously described in the 
Standard Model. In a simple extension in which new physics 
enters dominantly through $Z^0$ penguins with a CP-violating phase, 
only certain $B\to \pi K$ modes are affected by new physics. The 
$B\to \pi \pi$ data can then be described entirely within the Standard Model 
but with values of hadronic parameters that reflect large non-factorizable 
contributions. Using the $SU(3)$ flavour symmetry and plausible dynamical 
assumptions, we can then use the $B\to\pi\pi$ decays to fix the hadronic 
part of the $B \to \pi K$ system and make predictions for various observables 
in the $B_d \to \pi^{\mp} K^{\pm}$ and $B^{\pm} \to \pi^{\pm} K$ decays that 
are practically unaffected by electroweak penguins. The data on the  
$B^{\pm} \to \pi^0 K^{\pm}$ and $B_d \to \pi^0 K$ modes allow us then 
to determine the electroweak penguin component which differs from the 
Standard Model one, in particular through a large additional CP-violating 
phase. The implications for rare $K$ and $B$ decays are spectacular. In 
particular, the rate for $\klpn$ is enhanced by one order of magnitude, the 
branching ratios for $B_{d,s}\to \mu^+ \mu^-$ by a factor of five, and 
$\mathrm{BR}(K_{\rm L}\to \pi^0 e^+e^-, \pi^0 \mu^+ \mu^-)$ 
%$\mathrm{BR}(K_{\rm L}\to \pi^0 \mu^+ \mu^-)$ 
by factors of three. 
%We give a list of predictions for CP asymmetries in $B\to \pi \pi$ and 
%$B\to \pi K$ systems and for the rates of rare decays. 
%The overall agreement with the data is very good.
}

\normalsize\baselineskip=15pt

\section{Introduction}
The Standard Model (SM) for strong and electroweak interactions of quarks and
leptons gives a very satisfactory description of the observed phenomena down to
the short-distance scales of ${\cal O}(10^{-18})\,m$ or equivalently 
the scale of the top-quark mass. Exceptions are the non-vanishing neutrino 
masses, possibly related to scales of ${\cal O}(10^{-28})\, m$
through the see-saw mechanism, the observed matter--antimatter asymmetry
of the Universe, possibly related to even lower scales, and the issue of 
the low Higgs mass, which is related to the so-called naturalness
problem. One should, however, emphasize that, whereas the SM gauge sector of
the electroweak interactions has been tested to a very high precision in the 
1990s, studies of the flavour-changing interactions of quarks and leptons --
in particular those involving CP-violating transitions -- did not yet reach 
this precision and we should be prepared for surprises. Finally, the
non-perturbative part of QCD has to be put under much better control.

In this context, the present studies of non-leptonic two-body $B$ decays 
and of rare $K$ and $B$ decays are very important as they will teach us 
both about the non-perturbative aspects of QCD and about the perturbative 
electroweak physics at very short distances. For the analysis of these
modes, it is essential to have a strategy available that could clearly 
distinguish between non-perturbative QCD effects and short-distance 
electroweak effects. A strategy that in the case of deviations from 
the SM expectations would allow us transparently to identify a possible 
necessity for modifications in our understanding of hadronic effects and 
for a change of the SM model picture of electroweak flavour-changing 
interactions at short-distance scales.

In \cite{BFRS-PRL,BFRS-BIG}, we have developed a strategy
that allows us to address these questions in a systematic manner. It
encompasses non-leptonic $B$ and $K$ decays and rare $K$ and $B$ decays 
but has been at present used primarily for the analysis of 
$B\to\pi\pi$ and $B\to\pi K$ systems and rare $K$ and $B$ decays. The 
purpose of this note is to summarize the basic ingredients of our strategy 
and to list the most important results. A detailed update of the 
analyses in \cite{BFRS-PRL,BFRS-BIG} has recently been presented 
in \cite{BFRS-UPDATE}. The outline is as follows: in Section 2, we 
discuss the general aspects of the strategy proposed in 
\cite{BFRS-PRL,BFRS-BIG}. Sections 3 and 4 are devoted to the $B\to\pi\pi$ 
and $B\to\pi K$ systems, respectively. We consider rare $K$ and $B$ decays 
in Section 5, and give a short outlook in Section 6.

\section{Basic Strategy}
In order to illustrate our strategy in explicit terms, we shall 
consider a simple extension of the SM in which new physics (NP) 
enters dominantly through enhanced $Z^0$ penguins involving a 
CP-violating weak phase. As we will see below, this choice is 
dictated by the pattern of the data on the $B\to\pi K$ observables 
and the great predictivity of this scenario. It was first considered 
in \cite{Buras:1998ed,BRS,Buras:1999da} to study correlations between 
rare $K$ decays and the ratio $\epe$ measuring direct CP violation
in the neutral kaon system, and was generalized to rare $B$ decays in 
\cite{Buchalla:2000sk}. Here we extend these considerations to non-leptonic
$B$-meson decays, which allows us to confront this extension of the SM with
many more experimental results. Our strategy consists of three interrelated 
steps, and has the following logical structure:

\vspace*{0.3truecm}

\noindent
{\bf Step 1:}

\noindent
Since $B\to\pi\pi$ decays and the usual analysis of the unitarity 
triangle (UT) are only insignificantly affected by electroweak (EW) 
penguins, the $B\to\pi\pi$ system can be 
described as in the SM. Employing the $SU(2)$ isospin flavour symmetry
of strong interactions and the information on $\gamma$ from the
UT fits, we may extract the relevant hadronic parameters, and find large 
non-factorizable contributions, which are in particular reflected by 
large CP-conserving strong phases. Having these parameters at hand, we 
may then also predict the direct and mixing-induced CP asymmetries
of the $B_d\to\pi^0\pi^0$ channel. A future measurement of one of these
observables allows a determination of $\gamma$.

\vspace*{0.3truecm}

\noindent
{\bf Step 2:}

\noindent
If we use the $SU(3)$ flavour symmetry and plausible dynamical 
assumptions, we may determine the hadronic $B\to\pi K$ parameters 
through the $B\to\pi\pi$ analysis, and may calculate the $B\to\pi K$ 
observables in the SM. Interestingly, we find agreement with the pattern 
of the $B$-factory data for those observables where EW penguins play only 
a minor r\^ole. On the other hand, the observables receiving significant
EW penguin contributions do {\it not} agree with the experimental picture, 
thereby suggesting NP in the EW penguin sector. Indeed, a detailed analysis
shows \cite{BFRS-PRL,BFRS-BIG,BFRS-UPDATE}
that we may describe all the currently available data through moderately 
enhanced EW penguins with a large CP-violating NP phase around $-90^\circ$.
A future test of this scenario will be provided by the CP-violating 
$B_d\to \pi^0 K_{\rm S}$ observables, which we may predict. Moreover, 
we may obtain valuable insights into $SU(3)$-breaking effects, which 
support our working assumptions, and may also determine the UT angle 
$\gamma$, in remarkable agreement with the well-known UT fits.  

\vspace*{0.3truecm}

\noindent
{\bf Step 3:}

\noindent
In turn, the modified EW penguins with their large CP-violating 
NP phase have important implications for rare $K$ and $B$ decays.
Interestingly, several predictions differ significantly from the SM 
expectations and should easily be identified once the data improve. 
Similarly, we may explore specific NP patterns in other non-leptonic 
$B$ decays such as $B_d\to\phi K_{\rm S}$.

\begin{figure}
\begin{center}
\includegraphics[width=11.5cm,angle=-90]{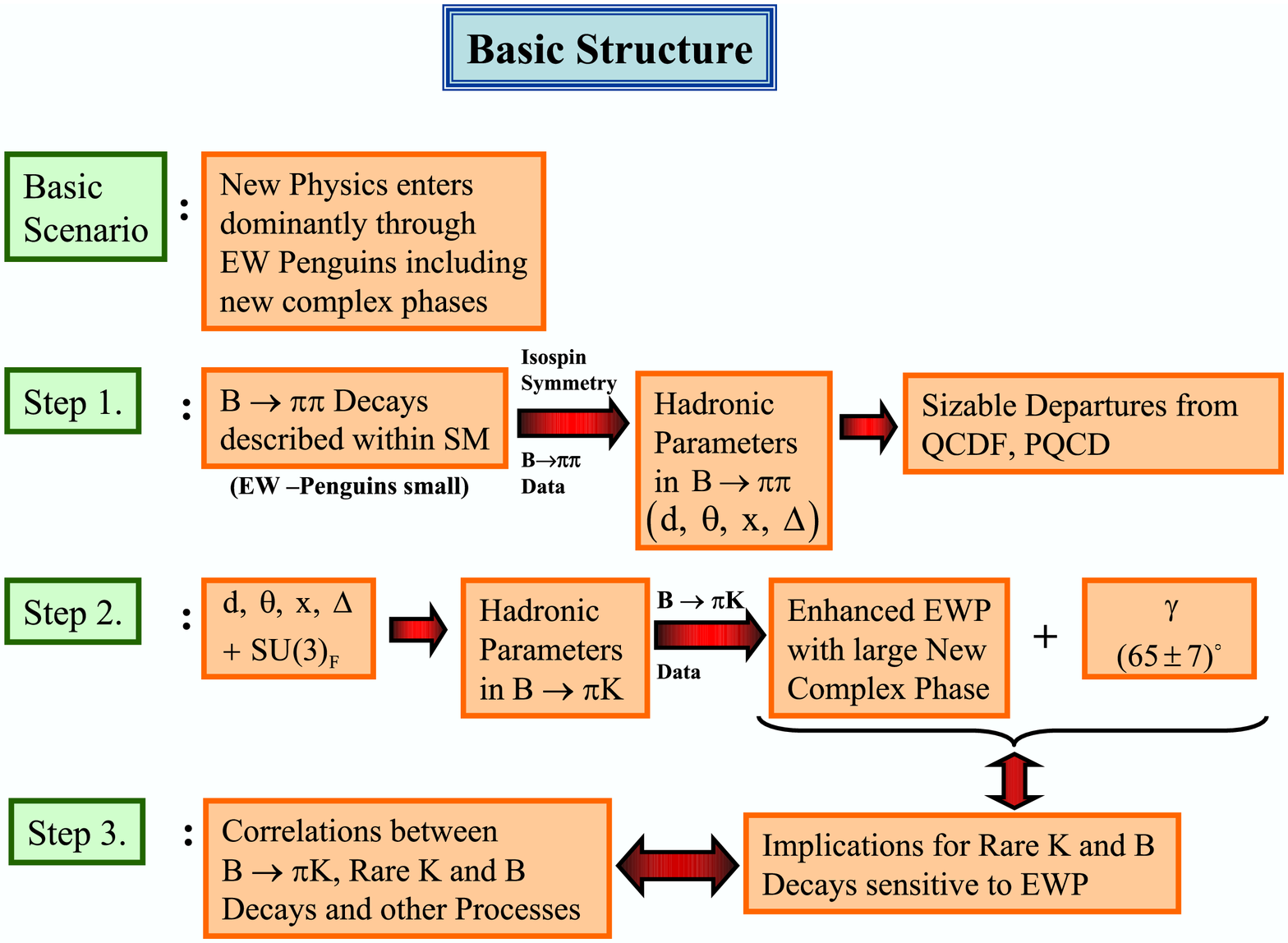}
\end{center}
\caption{Outline of our strategy}\label{chart}
\end{figure}

\vspace*{0.6truecm}

A chart of the three steps in question is given in Fig.\ \ref{chart}.
Before going into the details it is important to emphasize that our strategy is
valid both in the SM and all SM extensions in which NP enters
predominantly through the EW penguin sector. This means that even if
the presently observed deviations from the SM in the $B\to\pi K$ sector would
diminish with improved data, our strategy would still be useful in correlating
the phenomena in $B\to\pi\pi$, $B\to\pi K$ and rare $K$ and $B$ decays within
the SM. If, on the other hand, the observed deviations from the SM in
$B\to\pi\pi$ decays would not be attributed to the modification in hadron
dynamics but to NP contributions, our approach should be properly
generalized.

\boldmath
\section{$B\to\pi\pi$ decays}
\unboldmath
The central quantities for our analysis of the $B\to\pi\pi$ decays are the
ratios
\begin{eqnarray}
R_{+-}^{\pi\pi}&\equiv&2\left[\frac{\mbox{BR}(B^+\to\pi^+\pi^0)
+\mbox{BR}(B^-\to\pi^-\pi^0)}{\mbox{BR}(B_d^0\to\pi^+\pi^-)
+\mbox{BR}(\bar B_d^0\to\pi^+\pi^-)}\right]
\frac{\tau_{B^0_d}}{\tau_{B^+}}\label{Rpm-def}\\
R_{00}^{\pi\pi}&\equiv&2\left[\frac{\mbox{BR}(B_d^0\to\pi^0\pi^0)+
\mbox{BR}(\bar B_d^0\to\pi^0\pi^0)}{\mbox{BR}(B_d^0\to\pi^+\pi^-)+
\mbox{BR}(\bar B_d^0\to\pi^+\pi^-)}\right]\label{R00-def}
\end{eqnarray}
of the CP-averaged $B\to\pi\pi$ branching ratios, and the CP-violating 
observables provided by the time-dependent rate asymmetry 
\begin{eqnarray}
\lefteqn{\frac{\Gamma(B^0_d(t)\to \pi^+\pi^-)-\Gamma(\bar B^0_d(t)\to 
\pi^+\pi^-)}{\Gamma(B^0_d(t)\to \pi^+\pi^-)+\Gamma(\bar B^0_d(t)\to 
\pi^+\pi^-)}}\nonumber\\
&&={\cal A}_{\rm CP}^{\rm dir}(B_d\to \pi^+\pi^-)\cos(\Delta M_d t)+
{\cal A}_{\rm CP}^{\rm mix}(B_d\to \pi^+\pi^-)
\sin(\Delta M_d t)\label{rate-asym}\,.
\end{eqnarray}
The current status of the $B\to\pi\pi$ data together with the relevant
references can be found in Table \ref{tab:Bpipi-input}.
\begin{table}
%\vspace{0.4cm}
\begin{center}
%\vspace*{2mm}
\begin{tabular}{|c||c|c|}
\hline
Quantity & Input & Exp.\ reference
\\ \hline
 \hline
$\mbox{BR}(B^\pm\to\pi^\pm\pi^0)/10^{-6}$ & $5.5\pm0.6$ &
\cite{BaBar-Bpi0pi0, Chao:2003ue}
\\ \hline 
$\mbox{BR}(B_d\to\pi^+\pi^-)/10^{-6}$ & $4.6\pm0.4$ & 
\cite{Aubert:2002jb, Chao:2003ue} \\ \hline
$\mbox{BR}(B_d\to\pi^0\pi^0)/10^{-6}$ & $1.51\pm0.28$ & 
\cite{BaBar-Bpi0pi0, Belle-Bpi0pi0-new}
\\ \hline
 \hline
$R_{+-}^{\pi\pi}$ & $2.20\pm0.31$ & 
\\ \hline
$R_{00}^{\pi\pi}$ & $0.67\pm0.14$ & 
\\ \hline
 \hline
${\cal A}_{\rm CP}^{\rm dir}(B_d\to \pi^+\pi^-)$ & $-0.37\pm 0.11$ & 
\cite{BaBar-CP-Bpipi, Belle-CP-Bpipi}
\\ \hline
${\cal A}_{\rm CP}^{\rm mix}(B_d\to \pi^+\pi^-)$ & $+0.61\pm0.14$ & 
\cite{BaBar-CP-Bpipi, Belle-CP-Bpipi}
\\ \hline
\end{tabular}
\end{center}
\caption[]{The current status of the $B\to\pi\pi$ input data for our
strategy, with averages taken from \cite{HFAG}. For the 
evaluation of $R_{+-}^{\pi\pi}$, we have used the life-time ratio
$\tau_{B^+}/\tau_{B^0_d}=1.086\pm0.017$ \cite{PDG}.}\label{tab:Bpipi-input}
\end{table}
The so-called ``$B\to\pi\pi$ puzzle'' is reflected in a surprisingly 
large value of $\mbox{BR}(B_d\to\pi^0\pi^0)$ and a somewhat small
value of $\mbox{BR}(B_d\to\pi^+\pi^-)$, which results in
large values of both $R_{00}^{\pi\pi}$ and $R_{+-}^{\pi\pi}$. For instance,
the central values calculated within QCD factorization (QCDF) \cite{BBNS}
give $R_{00}^{\pi\pi}=0.07$ and $R_{+-}^{\pi\pi}=1.24$ \cite{Be-Ne}, although in the
scenario ``S4'' of \cite{Be-Ne} values 0.2 and 2.0, respectively, can
be obtained. As already pointed out in \cite{BFRS-PRL}, these data
indicate important non-factorizable contributions rather than NP
effects, and can be perfectly accommodated in the SM. The same applies
to the NP scenario considered in \cite{BFRS-PRL,BFRS-BIG}, in which
the EW penguin contributions to $B\to\pi\pi$ are marginal.

In order to address this issue in explicit terms, we use the isospin
symmetry to find
\begin{eqnarray}
\sqrt{2}A(B^+\to\pi^+\pi^0)&=&-[\tilde T+\tilde C] = 
-[T+C]\label{B+pi+pi0}\\
A(B^0_d\to\pi^+\pi^-)&=&-[\tilde T + P]\label{Bdpi+pi-}\\
\sqrt{2}A(B^0_d\to\pi^0\pi^0)&=&-[\tilde C - P].\label{Bdpi0pi0}
\end{eqnarray}
The individual amplitudes of (\ref{B+pi+pi0})--(\ref{Bdpi0pi0}) can
be expressed as
\begin{eqnarray}
P&=&\lambda^3 A({\cal P}_t-{\cal P}_c)\equiv\lambda^3 A {\cal P}_{tc}
\label{P-def}\\
\tilde T &=&\lambda^3 A R_b e^{i\gamma}\left[{\cal T}-\left({\cal 
P}_{tu}-{\cal E}\right)\right]\label{T-tilde}\\
\tilde C &=&\lambda^3 A R_b e^{i\gamma}\left[{\cal C}+\left({\cal P}_{tu}-
{\cal E}\right)\right],\label{C-tilde}
\end{eqnarray}
where 
\begin{equation}
\lambda\equiv|V_{us}|=0.2240\pm 0.0036, \quad
A\equiv |V_{cb}|/\lambda^2=0.83\pm0.02 
\end{equation}
are the usual parameters  in the Wolfenstein expansion of the 
Cabibbo--Kobayashi--Maskawa (CKM) matrix \cite{WO,BLO},
\begin{equation}\label{Rb-def}
R_b\equiv \sqrt{\bar\rho^2 + \bar\eta^2}=
\left(1-\frac{\lambda^2}{2}\right)
\frac{1}{\lambda}\left|\frac{V_{ub}}{V_{cb}}\right|=
0.37\pm0.04
\end{equation}
measures one side of the UT, the ${\cal P}_q$ describe the strong 
amplitudes of QCD penguins with internal $q$-quark exchanges 
($q\in\{t,c,u\}$), including annihilation and exchange penguins,
while ${\cal T}$ and ${\cal C}$ are the strong amplitudes of 
colour-allowed and colour-suppressed tree-diagram-like topologies, 
respectively, and ${\cal E}$ denotes the strong amplitude of an 
exchange topology. The amplitudes $\tilde T$ and $\tilde C$ differ from
\begin{equation}
T =\lambda^3 A R_b e^{i\gamma}{\cal T}, \quad
C =\lambda^3 A R_b e^{i\gamma}{\cal C}
\end{equation}
through the $({\cal P}_{tu}-{\cal E})$ pieces, which may play
an important r\^ole \cite{PAP0}. Note that these terms contain 
also the ``GIM penguins'' with internal up-quark exchanges, 
whereas their ``charming penguin'' counterparts enter in $P$ 
through ${\cal P}_{c}$, as can be seen in (\ref{P-def}) 
\cite{PAP0,c-pen,BFM,BPRS}. 

In order to characterize the dynamics of the $B\to\pi\pi$ system, we
introduce four hadronic parameters $d$, $\theta$, $x$ and $\Delta$ 
through
\begin{equation}\label{d-theta-def}
d e^{i\theta}=
-\left|\frac{P}{\tilde T}\right|e^{i(\delta_P-\delta_{\tilde T})}\,,
\quad
x e^{i\Delta}=
\left|\frac{\tilde C}{\tilde T}\right|e^{i(\delta_{\tilde C}-
\delta_{\tilde T})}\,,
\end{equation}
with $\delta_i$ being strong phases.
Using this parametrization, we have
\begin{equation}\label{Rpipi-gen}
R_{+-}^{\pi\pi}=F_1(d,\theta,x,\Delta;\gamma), \quad
R_{00}^{\pi\pi}=F_2(d,\theta,x,\Delta;\gamma),
\end{equation}
\begin{equation}\label{CP-Bpipi-gen}
{\cal A}_{\rm CP}^{\rm dir}(B_d\to \pi^+\pi^-)=
G_1(d,\theta;\gamma), \quad
{\cal A}_{\rm CP}^{\rm mix}(B_d\to \pi^+\pi^-)=
G_2(d,\theta;\gamma,\phi_d),
\end{equation}
with explicit expressions for $F_1$, $F_2$, $G_1$ and $G_2$
given in \cite {BFRS-BIG}. Taking then as the input
\begin{equation}\gamma=(65\pm7)^\circ\,,\quad
\phi_d=2\beta=\left(46.5^{+3.2}_{-3.0}\right)^\circ
\end{equation}
and the data for $R_{+-}^{\pi\pi}$, $R_{00}^{\pi\pi}$,
${\cal A}_{\rm CP}^{\rm dir}$ and ${\cal A}_{\rm CP}^{\rm mix}$
of Table 1, we obtain a set of four equations with the four unknowns
$d$, $\theta$, $x$ and $\Delta$. Interestingly, as demonstrated in
\cite{BFRS-BIG,BFRS-UPDATE}, a unique solution for these
parameters can be found:\footnote{In these results, also the tiny
EW penguin contributions to the $B\to\pi\pi$ decays are included 
\cite{BFRS-UPDATE}.}
\begin{equation} \label{dthetaxdelta}
d=0.51^{+0.26}_{-0.20}, \quad \theta=\left(140^{+14}_{-18}\right)^\circ, \quad 
x=1.15^{+0.18}_{-0.16}, \quad \Delta=-\left(59^{+19}_{-26}\right)^\circ.
\end{equation}
The large values of the strong phases $\theta$ and $\Delta$ and the
large values of $d$ and $x$ signal departures from the picture of
QCDF. Going back to (\ref{T-tilde}) and (\ref{C-tilde}) we observe
that these effects can be attributed to the enhancement of the
$({\mathcal{P}}_{tu}-{\cal E})$ terms that in turn suppresses 
${\tilde T}$ and enhances ${\tilde C}$.
In this manner, $\mbox{BR}(B_d\to\pi^+\pi^-)$ and
$\mbox{BR}(B_d\to\pi^0\pi^0)$ are suppressed and enhanced,
respectively. Moreover, also a sizeable deviation of the
``colour-suppression'' parameter 
$a_2^{\pi\pi}e^{i\Delta_2^{\pi\pi}}\equiv {\cal C}_{\pi\pi}/{\cal T}_{\pi\pi}$
from its naive value around 0.25 is suggested by the data, with 
$0.5\lsim a_2^{\pi\pi} \lsim 0.7$ and $\Delta_2^{\pi\pi}\sim 290^\circ$
\cite{FR-I}.

With the hadronic parameters at hand, we can predict the direct
and mixing-induced CP asymmetries of the $B_d\to\pi^0\pi^0$ channel.
These predictions, while still subject to large uncertainties, have
been confirmed by the most recent data. On the other hand, as
illustrated in \cite{BFRS-BIG}, a future precise measurement of
${\cal A}_{\rm CP}^{\rm dir}(B_d\to \pi^0\pi^0)$ or
${\cal A}_{\rm CP}^{\rm mix}(B_d\to \pi^0\pi^0)$ allows a theoretically
clean determination of $\gamma$.

The large non-factorizable effects found in 
\cite{BFRS-PRL} have been discussed at length in
\cite{BFRS-BIG,BFRS-UPDATE}, and have been confirmed in 
\cite{BPRS, ALP-Bpipi,CGRS,He:2004ck,Bauer:2004ck,Feldmann:2004mg}. 
For the following sections, 
the most important outcome of this analysis are the values of the 
hadronic parameters $d$, $\theta$, $x$ and $\Delta$ in (\ref{dthetaxdelta}).
These quantities will allow us -- with the help of the $SU(3)$ flavour 
symmetry -- to determine the corresponding hadronic parameters of the 
$B\to\pi K$ system.

\boldmath
\section{$B\to\pi K$ decays}
\unboldmath
The key observables for our discussion are the following ratios:
\begin{eqnarray}
R&\equiv&\left[\frac{\mbox{BR}(B_d^0\to\pi^- K^+)+
\mbox{BR}(\bar B_d^0\to\pi^+ K^-)}{\mbox{BR}(B^+\to\pi^+ K^0)+
\mbox{BR}(B^-\to\pi^- \bar K^0)}
\right]\frac{\tau_{B^+}}{\tau_{B^0_d}}\label{R-def}\\
R_{\rm c}&\equiv&2\left[\frac{\mbox{BR}(B^+\to\pi^0K^+)+
\mbox{BR}(B^-\to\pi^0K^-)}{\mbox{BR}(B^+\to\pi^+ K^0)+
\mbox{BR}(B^-\to\pi^- \bar K^0)}\right]\label{Rc-def}\\
R_{\rm n}&\equiv&\frac{1}{2}\left[
\frac{\mbox{BR}(B_d^0\to\pi^- K^+)+
\mbox{BR}(\bar B_d^0\to\pi^+ K^-)}{\mbox{BR}(B_d^0\to\pi^0K^0)+
\mbox{BR}(\bar B_d^0\to\pi^0\bar K^0)}\right],\label{Rn-def}
\end{eqnarray}
where the current status of the relevant branching ratios and the
corresponding values of the $R_i$ is summarized in 
Table \ref{tab:BpiK-input1}.
\begin{table}
%\vspace{0.4cm}
\begin{center}
%\vspace*{2mm}
\begin{tabular}{|c||c|c|}
\hline
Quantity & Data & Exp. reference
\\ \hline
 \hline
$\mbox{BR}(B_d\to\pi^\mp K^\pm)/10^{-6}$ & $18.2\pm0.8$ & 
\cite{Aubert:2002jb, Chao:2003ue}
\\ \hline
$\mbox{BR}(B^\pm\to\pi^\pm K)/10^{-6}$ & $24.1\pm1.3$ &
\cite{BaBar-BdKK-obs, Chao:2003ue}
\\ \hline
$\mbox{BR}(B^\pm\to\pi^0K^\pm)/10^{-6}$ & $12.1\pm0.8$ & 
\cite{BaBar-Bpi0pi0, Chao:2003ue}
\\ \hline
$\mbox{BR}(B_d\to\pi^0K)/10^{-6}$ & $11.5\pm1.0$ & 
\cite{BaBar-BK0pi0, Chao:2003ue}
\\ \hline
 \hline
$R$ & $0.82\pm0.06$ & $0.91\pm0.07$
\\ \hline
$R_{\rm c}$ & $1.00\pm0.08$ & $1.17\pm0.12$
\\ \hline
$R_{\rm n}$ & $0.79\pm0.08$ & $0.76\pm0.10$
\\ \hline
\end{tabular}
\end{center}
\caption[]{The current status of the CP-averaged $B\to\pi K$ 
branching ratios, with averages taken from \cite{HFAG}. We also give the
values of the ratios $R$, $R_{\rm c}$ and $R_{\rm n}$ introduced
in (\ref{R-def}), (\ref{Rc-def}) and (\ref{Rn-def}), where  
$R$ refers again to $\tau_{B^+}/\tau_{B^0_d}=1.086\pm0.017$ 
\cite{PDG}. In the last column we also give the values of
$R_i$ at the time of the analyses in \cite{BFRS-PRL,BFRS-BIG}.
}\label{tab:BpiK-input1}
\end{table}
The so-called ``$B\to\pi K$ puzzle'', which was already pointed out in
\cite{BF-neutral2}, is reflected in the small value of
$R_{\rm n}$ that is significantly lower than $R_{\rm c}$.
We will return to this below.

In order to analyze this issue, we neglect for simplicity the
colour-suppressed EW penguins and use the $SU(3)$ flavour symmetry 
to find:
\begin{eqnarray}
A(B^0_d\to\pi^-K^+)&=&P'\left[1-re^{i\delta}e^{i\gamma}
\right]\label{Bdpi-K+}\\
A(B^{\pm}_d\to\pi^{\pm} K^0)&=&-P'\label{B+pi-K0}\\
\sqrt{2}A(B^+\to\pi^0K^+)&=&P'\left[1-\left(e^{i\gamma}-qe^{i\phi}\right)
r_{\rm c}e^{i\delta_{\rm c}}\right]\label{B+pi0K+-SM}\\
\sqrt{2}A(B^0_d\to\pi^0K^0)&=&-P'\left[1+\rho_{\rm n}e^{i\theta_{\rm n}}
e^{i\gamma}-qe^{i\phi}r_{\rm c}e^{i\delta_{\rm c}}\right].\label{B0pi0K0-SM}
\end{eqnarray}
Here, $P'$ is a QCD penguin amplitude that does not concern us as it cancels 
in the ratios $R_i$ and in the CP asymmetries considered. The parameters $r$, 
$\delta$, $\rho_{\rm n}$, $\theta_{\rm n}$, $r_{\rm c}$ and $\delta_{\rm c}$ 
are of hadronic origin. If they were considered as completely free, the 
predictive power of (\ref{Bdpi-K+})--(\ref{B0pi0K0-SM}) would be rather low. 
Fortunately, using the $SU(3)$ flavor symmetry, they can be related to the 
parameters $d$, $\theta$, $x$ and $\Delta$ that we have determined in the 
previous section; explicit expressions can be found in \cite{BFRS-BIG}. In 
this manner, the values of $r$, $\delta$, $\rho_{\rm n}$, $\theta_{\rm n}$, 
$r_{\rm c}$ and $\delta_{\rm c}$ can be found.
%, subject to non-factorizable $SU(3)$-breaking corrections and 
%the uncertainties in the $B\to \pi \pi$ data. 
The specific numerical values for these parameters are not of particular 
interest here and can be found in \cite{BFRS-UPDATE}. It suffices to say 
that they also signal large non-factorizable hadronic effects. 

\begin{figure}
%\vspace*{0.3truecm}
\begin{center}
\includegraphics[width=13cm]{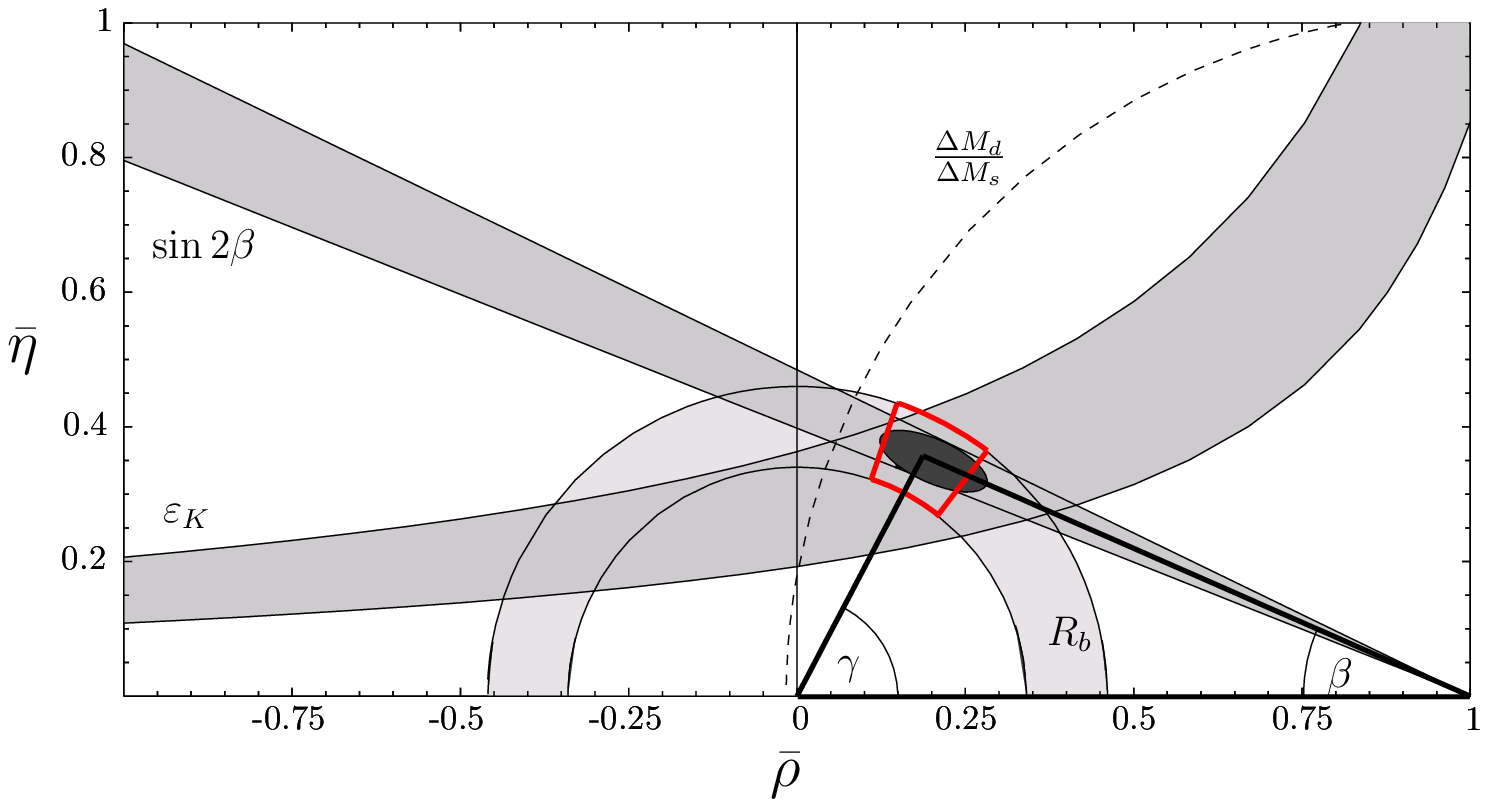}
\end{center}
%\vspace*{-0.5truecm}
\caption{Illustration of the value of $\gamma$ following from the 
CP-violating $B_d\to\pi^+\pi^-$ observables and the data for the
$B_d\to\pi^\mp K^\pm$ decays in the $\bar\rho$--$\bar \eta$ plane 
and comparison with the other constraints for the UT, as discussed
in \cite{BSU}. The shaded dark ellipse is the result of the UT fit,
while the quadrangle corresponds to the second value of
$\left.\gamma\right|_{\rm BR}$ in (\ref{gamma}) and the $R_b$ 
constraint.}\label{fig:UT}
\end{figure}

The most important recent experimental result concerning the $B\to\pi K$
system is the observation of direct CP violation in $B_d\to\pi^\mp K^\pm$
decays \cite{BaBar-CP-dir-obs,Belle-CP-dir-obs}. This phenomenon is 
described by the following rate asymmetry:
\begin{equation}
{\cal A}_{\rm CP}^{\rm dir}(B_d\to\pi^\mp K^\pm)\equiv
\frac{\mbox{BR}(B^0_d\to\pi^-K^+)-
\mbox{BR}(\bar B^0_d\to\pi^+K^-)}{\mbox{BR}(B^0_d\to\pi^-K^+)+
\mbox{BR}(\bar B^0_d\to\pi^+K^-)}=+0.113\pm 0.019,
\end{equation}
where the numerical value is the average compiled in \cite{HFAG}. Using 
the values of $r$ and $\delta$ as determined above and (\ref{Bdpi-K+}), 
we obtain 
${\cal A}_{\rm CP}^{\rm dir}(B_d\to\pi^\mp K^\pm)=+0.127^{+0.102}_{-0.066}$,
which is in nice accordance with the experimental result. In fact, we 
argued in our previous analysis \cite{BFRS-BIG}, which implied
{\small $+0.140^{+0.139}_{-0.087}$} and was confronted with 
the experimental average $+0.095\pm0.028$, that this CP asymmetry 
should go up. Following the lines of \cite{RF-Bpipi,Fl-Ma}, 
we may determine the angle $\gamma$ of the UT by complementing 
the CP-violating $B_d\to\pi^+\pi^-$ observables with either the 
ratio of the CP-averaged branching ratios $\mbox{BR}(B_d\to\pi^\mp K^\pm)$ 
and $\mbox{BR}(B_d\to\pi^+\pi^-)$ or the direct CP-asymmetry 
${\cal A}_{\rm CP}^{\rm dir}(B_d\to\pi^\mp K^\pm)$. 
These avenues, where the latter is theoretically more favourable, 
yield the following results: 
\begin{equation}
\label{gamma}
\left.\gamma\right|_{\rm BR}=
\left(63.3\,^{+7.7}_{-11.1}\right)^\circ, \quad
\left.\gamma\right|_{{\cal A}_{\rm CP}^{\rm dir}}= 
\left(66.6\,^{+11.0}_{-11.1}\right)^\circ, 
\end{equation}
which are nicely consistent with each other. Moreover, these ranges
are in remarkable accordance with the SM picture, as can be seen in 
Fig.\ \ref{fig:UT}, where we compare these values of $\gamma$ with 
the UT fit performed in \cite{BSU}. A similar extraction of $\gamma$
can be found in \cite{Wu:2004xx}.

%If we take also the experimental
%value of $R_b$ into account, we obtain
%\begin{equation}
%\left.\alpha\right|_{\rm BR}=\left(95.0\,^{+12.2}_{-8.2}\right)^\circ, \quad
%\left.\alpha\right|_{{\cal A}_{\rm CP}^{\rm dir}}=
%\left(91.7\,^{+12.0}_{-11.0}\right)^\circ,
%\label{alpha-determ}\end{equation}
%\begin{equation}
%\left.\beta\right|_{\rm BR}=\left(21.6\,^{+2.6}_{-2.7}\right)^\circ, \quad
%\left.\beta\right|_{{\cal A}_{\rm CP}^{\rm dir}}=
%\left(21.7\,^{+2.5}_{-2.6}\right)^\circ.
%\label{beta-determ}\end{equation}

Apart from ${\cal A}_{\rm CP}^{\rm dir}(B_d\to\pi^\mp K^\pm)$,
two observables are left that are only marginally affected by
EW penguins: the ratio $R$ introduced in (\ref{R-def}) and
the direct CP asymmetry of $B^\pm\to\pi^\pm K$ modes. These observables
may be affected by another hadronic parameter $\rho_{\rm c}
e^{i\theta_{\rm c}}$, which is expected to play a minor r\^ole and
was neglected in (\ref{B+pi-K0}) and (\ref{B+pi0K+-SM}). In this
approximation, the direct $B^\pm\to\pi^\pm K$ CP asymmetry vanishes --
in accordance with the experimental value of $+0.020\pm0.034$ -- and 
the new experimental value of $R=0.82\pm0.06$, which is on the lower 
side, can be converted into $\gamma\leq (64.9^{+4.8}_{-4.2})^\circ$ 
with the help of the bound derived in \cite{FM}. On the other hand,
if we use the values of $r$ and $\delta$ as determined above, we
obtain
\begin{equation}\label{R-pred}
R=0.943^{+0.028}_{-0.021},
\end{equation}
which is sizeably larger than the experimental value. The nice agreement 
of the data with our prediction of 
${\cal A}_{\rm CP}^{\rm dir}(B_d\to\pi^\mp K^\pm)$, which 
is independent of $\rho_{\rm c}$, suggests that this parameter is 
actually the origin of the deviation of $R$. In fact, as discussed
in detail in \cite{BFRS-UPDATE}, the emerging signal for 
$B^\pm\to K^\pm K$ decays, which provide direct access to
$\rho_{\rm c}$, shows that our value of $R$ in (\ref{R-pred}) 
is shifted towards the experimental value through this parameter,
thereby essentially resolving this small discrepancy. 

It is imporant to emphasize that we could accommodate all the 
$B\to\pi\pi$ and $B\to\pi K$ data so far nicely in the SM. Moreover,
as discussed in detail in \cite{BFRS-BIG,BFRS-UPDATE}, there are
also a couple of internal consistency checks of our working 
assumptions, which work impressively well within the current 
uncertainties.

Let us now turn our attention to (\ref{B+pi0K+-SM}) and (\ref{B0pi0K0-SM}). 
The only variables in these formulae that we did
not discuss so far are the parameters $q$ and $\phi$ that parametrize
the EW penguin sector. The fact that EW penguins cannot be neglected
here is related to the simple fact that a $\pi^0$ meson can be emitted
directly in these colour-allowed EW penguin topologies, while the
corresponding emission with the help of QCD penguins is 
colour-suppressed. In the SM, the parameters $q$ and $\phi$ can be 
determined with the help of the $SU(3)$ flavour symmetry of strong 
interactions \cite{NR}, yielding
\begin{equation}\label{q-SM}
q=0.69 \times\left[\frac{0.086}{|V_{ub}/V_{cb}|}\right], 
\qquad \phi=0^\circ.
\end{equation}
In this manner, predictions for $R_{\rm c}$ and $R_{\rm n}$ can be made
\cite{BFRS-PRL,BFRS-BIG}, which read as follows \cite{BFRS-UPDATE}: 
\begin{equation}\label{Rn-Rc-pred}
\left.R_{\rm c}\right|_{\rm SM}=1.14 \pm 0.05 ,\,\quad
\left. R_{\rm n}\right|_{\rm SM}=1.11^{+0.04}_{-0.05}\,.
\end{equation}
Comparing with the experimental results in Table~\ref{tab:BpiK-input1},
we observe that there is only a marginal discrepancy in the
case of $R_{\rm c}$, whereas the value of $R_{\rm n}$ in 
(\ref{Rn-Rc-pred}) is definetely too large. The ``$B\ \to \pi K$'' 
puzzle is seen here in explicit terms.

\begin{figure}
%\vspace*{0.3truecm}
\begin{center}
\includegraphics[width=12cm]{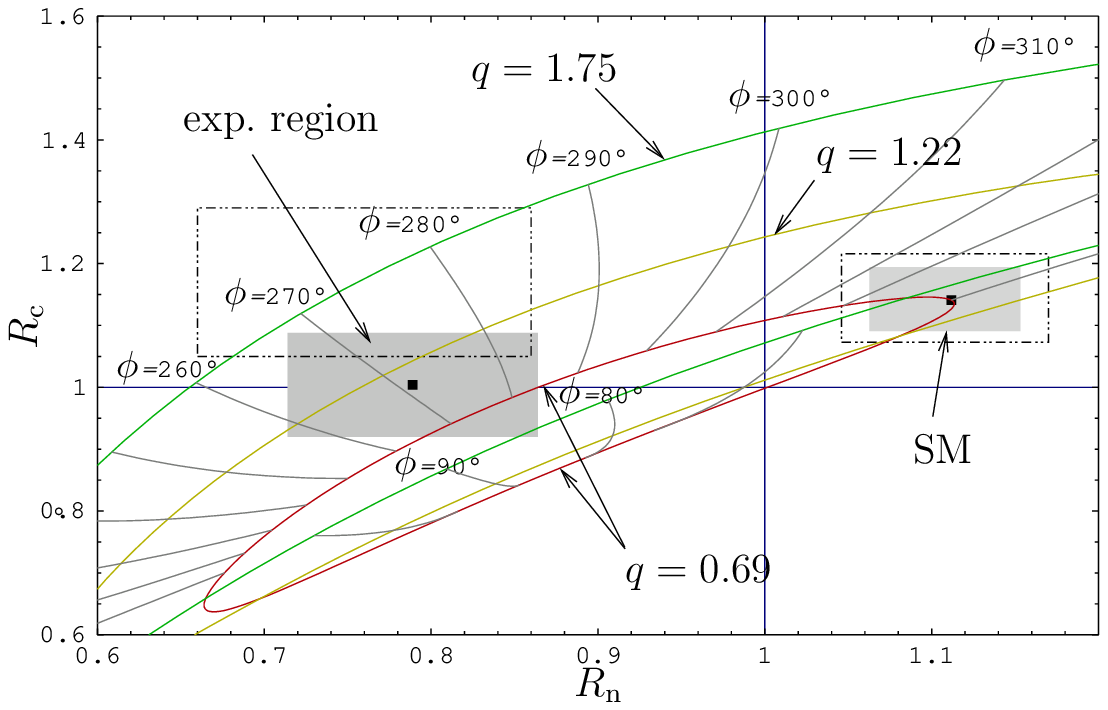}
\end{center}
\caption{The situation in the $R_{\rm n}$--$R_{\rm c}$ plane. We show contours
for values of $q=0.69$, $q=1.22$ and $q=1.75$, with
$\phi \in [0^\circ,360^\circ]$. 
The experimental ranges for $R_{\rm c}$ and $R_{\rm n}$ and those predicted 
in the SM are indicated in grey, the dashed lines serve as a reminder of
the corresponding ranges in \cite{BFRS-BIG}.\label{fig:Rn-Rc}}
\end{figure}

The disagreement of the SM with the data can be resolved in the
scenario of enhanced EW penguins carrying a non-vanishing phase
$\phi$. Indeed, using the measured values of $R_{\rm n}$ and 
$R_{\rm c}$, we find \cite{BFRS-UPDATE}:
\begin{equation}\label{q-det}
q = 1.08\,^{+0.81}_{-0.73} , \quad \, \phi =
-(88.8^{+13.7}_{-19.0})^\circ \,. 
\end{equation}
We observe that -- while $q$ is consistent with the SM estimate
within the errors -- the large phase $\phi$ is a spectacular signal
of possible NP contributions. It is useful to consider the 
$R_{\rm n}$--$R_{\rm c}$ plane, as we have done in Fig.~\ref{fig:Rn-Rc}. 
There we show contours corresponding to different values of $q$, and 
indicate the experimental and SM ranges. Following \cite{BFRS-BIG}, 
we choose the values of $q=0.69$, $1.22$ and $1.75$, where the latter 
reproduced the central values of $R_{\rm c}$ and $R_{\rm n}$ in our 
previous analysis \cite{BFRS-PRL,BFRS-BIG}. The central values for the 
SM prediction have hardly moved, while their uncertainties have been 
reduced a bit. On the other hand, the central experimental values of 
$R_{\rm c}$ and $R_{\rm n}$ have moved in such a way that $q$ decreased, 
while the weak phase $\phi$ remains around $-90^{\circ}$. 

We close this section with a list of predictions for the CP asymmetries
in the $B \to \pi \pi$ and $B\to \pi K$ systems, which are summarized
in Table~\ref{asymtab}.

\begin{table}[hbt]
%\vspace{0.4cm}
\begin{center}
%\vspace*{2mm}
\begin{tabular}{|c||c|c|}
\hline
  Quantity &   Our Prediction &  Experiment
 \\ \hline
  ${\cal A}_{\rm CP}^{\rm dir}(B_d\!\to\!\pi^0 \pi^0)$ 
 &   $-0.28^{+0.37}_{-0.21}$ \rule{0em}{1.05em} & $-0.28 \pm 0.39$ \\\hline
  ${\cal A}_{\rm CP}^{\rm mix}(B_d\!\to\!\pi^0 \pi^0)$ 
 &  $-0.63^{+0.45}_{-0.41}$ \rule{0em}{1.05em} &
 $-0.48_{-0.40}^{+0.48}$ \\\hline
${\cal A}_{\rm CP}^{\rm dir}(B_d\!\to\!\pi^{\mp} K^{\pm}) $ 
&   $0.127^{+0.102}_{-0.066}$ \rule{0em}{1.05em} &  $0.113\pm0.019$ \\
\hline
${\cal A}_{\rm CP}^{\rm dir}(B^\pm\!\to\!\pi^0 K^\pm) $ 
&   $0.10^{+0.25}_{-0.19}$ \rule{0em}{1.05em} &  $-0.04 \pm 0.04$ \\ \hline
  ${\cal A}_{\rm CP}^{\rm dir}(B_d\!\to\!\pi^0 K_{\rm S})$ 
&   $0.01^{+0.15}_{-0.18}$ \rule{0em}{1.05em} &
$0.09 \pm 0.14$ \\ \hline 
  ${\cal A}_{\rm CP}^{\rm mix}(B_d\!\to\!\pi^0 K_{\rm S})$ 
&  $-0.98^{+0.04}_{-0.02} $ \rule{0em}{1.05em} &  $-0.34^{+0.29}_{-0.27}$
\\ \hline
\end{tabular}
\caption{Compilation of our predictions for the CP-violating 
$B\to\pi\pi$, $\pi K$ asymmetries.}
\label{asymtab}
\end{center}
\end{table}

\boldmath
\section{Implications for Rare $K$ and $B$ Decays}
\unboldmath
The rates for rare $K$ and $B$ decays are sensitive functions
of the EW penguin contributions. In a simple scenario in
which NP enters the EW penguin sector predominantly through $Z^0$
penguins and the local operators in the effective Hamiltonians for
rare decays are the same as in the SM\footnote{See
 \cite{Barger:2004hn} for a discussion of the $B \to \pi K$ system in a
 slightly different scenario involving an additional $Z^{'}$ boson.},
there is a strict relation \cite{BFRS-I,BFRS-PRL,BFRS-BIG} between the
EW penguin effects  in the $B\to\pi K$ system and the corresponding
effects in rare  $K$ and $B$ decays. Denoting by $C$ the $Z^0$-penguin
function, we find 
\begin{equation}
C=|C|e^{i\theta_C}=2.35 \bar{q} e^{i\phi}-0.82, \qquad
\bar{q}=q\left|\frac{V_{ub}/V_{cb}}{0.086}\right| \,. 
\end{equation}
In turn, the functions $X$, $Y$, $Z$ that govern rare decays in the
scenarios in question become explicit functions of $q$ and $\phi$:
\begin{eqnarray}\label{Xfunc}
X=|X|e^{i\theta_X}&=&2.35 \bar{q}e^{i\phi}-0.09,
\\
Y=|Y|e^{i\theta_Y}&=&2.35 \bar{q}e^{i\phi}-0.64,
\\ \label{Zfunc}
Z=|Z|e^{i\theta_Z}&=&2.35 \bar{q}e^{i\phi}-0.94.
\end{eqnarray}

If the phase $\phi$ was zero (the case considered in \cite{BFRS-I}), 
the functions $X$, $Y$, $Z$ would remain to be real quantities as in 
the SM but the enhancement of $q$ would imply enhancements of 
$X$, $Y$, $Z$ as well. As in the scenario considered $X$, $Y$, $Z$ are 
not independent of one another, it is sufficient to constrain one of 
them from rare decays in order to see whether the enhancement of $q$ 
is consistent with the existing data on rare decays. It turns out that 
the data on inclusive $B\to X_s l^+l^-$ decays \cite{Kaneko:2002mr,Abe:2004sg} are presently most
powerful to constrain $X$, $Y$, $Z$, but due to significant
experimental errors and theoretical uncertainties these
bounds are only approximate. Typically, one finds $X$, $Y$, $Z$ to
be at most $2.2$ to be compared with $X\approx 1.5$, $Y\approx 1.0$, 
$Z\approx 0.7$, in the SM, thereby leaving still some room for 
NP contributions.

Now, $X$ governs decays with $\nu\bar\nu$ in the final state like 
$K\to\pi\nu\bar\nu$, $Y$
governs decays with $l^+l^-$ in the final state like 
$B_{s,d}\to\mu^+\mu^-$, $K_{\rm L}\to \pi^0 l^+l^-$ and
$B\to X_s l^+l^-$, while $Z$ is relevant for $K_{\rm L}\to \pi^0 l^+l^-$, 
$B\to X_s l^+l^-$ and in particular for $\epe$. 
In the situation with $\phi=0^\circ$ there is clearly room for sizable
departures in rare decay rates from the SM expectation \cite{BFRS-I},
but in particular the branching ratios for $K^+\to \pi^+\nu\bar\nu$ and 
$K_{\rm L}\to \pi^0\nu\bar\nu$ are only
enhanced by at most factors of $2$--$2.5$ as in other models with minimal
flavour violation (MFV) \cite{Buras:2000dm,D'Ambrosio:2002ex}
in which no new phases beyond the KM phase are present.

The situation changes drastically if $\phi$ is required to be
non-zero, in particular when its value is in the ball park
of $-90^\circ$ as found in the previous section. In this case, $X$, $Y$
and $Z$ become complex quantities, as seen in (\ref{Xfunc})--(\ref{Zfunc}), 
with the phases $\theta_i$ in the ball park of $-90^\circ$ 
\cite{BFRS-PRL,BFRS-BIG}:
\be\label{rXrYrZ}
 \theta_X= -(86\pm 12)^\circ, \qquad
 \theta_Y= -(100\pm 12)^\circ, \qquad 
 \theta_Z=-(108\pm 12)^\circ .
\ee
Actually, the data for the $B\to\pi K$ decays used in our first analysis 
\cite{BFRS-BIG} were such that $q=1.75^{+1.27}_{-0.99}$ and 
$\phi=-(85^{+11}_{-14})$ were required, implying
$|X|\approx|Y|\approx|Z|\approx4.3^{+3.0}_{-2.4}$, barely consistent
with the data. Choosing $|Y|=2.2$ as high as possible but still
consistent with the data on $B\to X_s l^+l^-$ , we found
\begin{equation} 
\label{qwithRD}
\bar{q}=0.92^{+0.07}_{-0.05}\,, \qquad \phi=-(85^{+11}_{-14})^{\circ}\,, 
\end{equation}
 which has been already taken into account in obtaining the values in 
(\ref{rXrYrZ}).
This in turn made us expect that the experimental values
$R_{\rm c}=1.17\pm0.12$ and $R_{\rm n}=0.76\pm0.10$ known at the time of the
analysis in \cite{BFRS-BIG} could change (see Fig.\ \ref{fig:Rn-Rc}) 
once the data improve. Indeed, our expectation \cite{BFRS-BIG}, 
\begin{equation}
R_{\rm c}=1.00^{+0.12}_{-0.08} \qquad R_{\rm n}=0.82^{+0.12}_{-0.11},
\end{equation}
has been confirmed by the most recent results in Table
\ref{tab:BpiK-input1}, 
making the overall description of the $B\to \pi \pi$, $B\to \pi K$ and 
rare decays within our approach significantly better with respect to our 
previous analysis.

There is a characteristic pattern of modifications of
branching ratios relative to the case of $\phi=0^\circ$ and 
$\theta_i=0^\circ$:
\begin{itemize}
\item  The branching ratios proportional to $|X|^2$, like 
$\mbox{BR}(B\to X_s\nu\bar\nu)$, and  $|Y|^2$,
like $\mbox{BR}(B_{d,s}\to \mu^+\mu^-)$, remain unchanged relative to
the case $\phi=0^\circ$, still exhibiting enhancements of
roughly $2$ and $5$, respectively, over the SM expectations.
\item  In CP-conserving transitions in which in addition to
top-like contributions also charm contribution plays some
r\^ole, the {\it constructive} interference between top and charm
contributions in the SM becomes {\it destructive}, thereby compensating 
for the enhancements of $X$, $Y$ and $Z$. In particular, 
$\mbox{BR}(K^+\to \pi^+\nu\bar\nu)$ turns out to be rather
close to the SM estimates, and the short-distance part of 
$\mbox{BR}(K_{\rm L}\to \mu^+\mu^-)$ is even smaller than in the SM.
\item Not surprisingly, the most spectacular impact of large
phases $\theta_i$ is seen in CP-violating quantities sensitive to EW
penguins.
\end{itemize}

In particular, one finds \cite{BFRS-PRL,BFRS-BIG}
\be
\frac{\mbox{BR}(\klpn)}{\mbox{BR}(\klpn)_{\rm SM}}=
\left|\frac{X}{X_{\rm SM}}\right|^2
\left[\frac{\sin(\beta-\theta_X)}{\sin(\beta)}\right]^2, 
\ee
with the two factors on the right-hand side in the ball park of $2$ and
$5$, respectively. Consequently, $\mbox{BR}(\klpn)$ can be enhanced over the
SM prediction even by an order of magnitude and is expected
to be roughly by a factor of $4$ larger than 
$\mbox{BR}(\kpn)$. In the SM and most MFV models the pattern is totally 
different with
$\mbox{BR}(\klpn)$ smaller than $\mbox{BR}(\kpn)$
by a factor of $2$--$3$ \cite{BSU,BBSIN,BF-MFV}. On the other hand a
recent analysis shows that a pattern of $BR(K \to \pi \nu \bar{\nu})$
expected in our NP 
scenario can be obtained in a general MSSM \cite{Buras:2004qb}.

We note that $\mbox{BR}(\klpn)$ is predicted to be rather close to 
its model-independent
upper bound~\cite{GRNR}
\be\label{BOUND}
\mbox{BR}(\klpn)\le 4.4\, \mbox{BR}(\kpn).
\ee
Moreover, another  spectacular 
implication of these findings is a strong violation of the relation 
\cite{BBSIN}
\be 
(\sin 2 \beta)_{\pi \nu\bar\nu}=(\sin 2 \beta)_{\psi K_{\rm S}},
\ee 
which is valid in the SM and any model with MFV. Indeed, we find
\cite{BFRS-PRL,BFRS-BIG}
\be
(\sin 2 \beta)_{\pi \nu\bar\nu}=\sin 2(\beta-\theta_X) =
-(0.69^{+0.23}_{-0.41}),
\ee
in striking disagreement with $(\sin 2 \beta)_{\psi K_{\rm S}}= 
0.725\pm0.037$.

Even if eventually the departures from the SM and MFV pictures
could turn out to
be smaller than estimated here, finding 
$\mbox{BR}(\klpn)$ to be larger than $\mbox{BR}(\kpn)$
would be a clear signal of new complex phases at work. For a more
general discussion of the $K \to \pi \nu \bar{\nu}$ decays beyond the
SM, see \cite{BSU}.

Similarly, as seen in Table~\ref{rare}, interesting enhancements are
found in $K_{\rm L}\to\pi^0 l^+l^-$ and the forward--backward CP asymmetry 
in $B\to X_sl^+l^-$ as discussed in \cite{BFRS-BIG}.

\begin{table}[hbt]
\begin{center}
\begin{tabular}{|c||c|c|c|}
\hline
 Decay & SM prediction &  Our scenario & 
Exp.\ bound \quad(\mbox{90\% {\rm C.L.}}) 
 \\ \hline
$K^+ \to \pi^+ \bar \nu \nu$  & $ (7.8 \pm 1.2) \cdot 10^{-11}$ 
&  $(7.5 \pm 2.1)\cdot 10^{-11}$ & $(14.7^{+13.0}_{-8.9})\cdot 10^{-11} $ 
\cite{Anisimovsky:2004hr} \\ \hline
 $K_{\rm L} \to \pi^0 \bar \nu \nu$   & $ (3.0 \pm 0.6)\cdot 10^{-11}$
&  $ (3.1\pm 1.0)\cdot 10^{-10}$ &  $ < 5.9 \cdot10^{-7}$ 
\cite{Alavi-Harati:1999hd} \\ \hline
 $K_{\rm L} \to \pi^0  e^+ e^-$ &  $ (3.7^{+1.1}_{-0.9})\cdot 10^{-11}$  &
 $(9.0\pm 1.6)\cdot 10^{-11}$  &   $<2.8\cdot 10^{-10}$ 
\cite{Alavi-Harati:2003mr} \\ \hline
$K_{\rm L} \to \pi^0  \mu^+ \mu^-$ &  $ (1.5 \pm 0.3)\cdot 10^{-11}$  &
 $(4.3\pm 0.7)\cdot 10^{-11}$  &   $<3.8\cdot 10^{-10}$ 
\cite{Alavi-Harati:2000hs} \\ \hline  
$B \to X_s\bar \nu \nu$  &  $(3.5\pm 0.5)\cdot 10^{-5}$  
&  $ \approx 7\cdot 10^{-5}$ &  $<6.4\cdot 10^{-4}$ 
\cite{Barate:2000rc} \\ \hline
  $B_s \to \mu^+ \mu^-$  &   $(3.42 \pm 0.53)\cdot 10^{-9}$ & 
  $\approx 17\cdot 10^{-9}$ &  $<5.0\cdot 10^{-7}$ 
\cite{Abazov:2004dj} \\ \hline
\end{tabular}
\end{center}
\caption[]{\small Predictions for various rare decays in the scenario
considered compared with the SM expectations and experimental
bounds. For a theoretical update on $K_{\rm L} \to \pi^0  e^+ e^-$ and 
a discussion of $K_{\rm L} \to \pi^0  \mu^+ \mu^-$, see \cite{Isidori:2004rb}.
\label{rare}}
\end{table}

As emphasized above, the new data on $B\to \pi K$ improved the overall
description of $B\to \pi\pi$, $B\to \pi K$ and rare decays within our
approach. But, as the rare decay constraint from $B\to X_s l^+l^- $
has already been taken into account in \cite{BFRS-BIG}, there is essentially no
impact of these new data on our predictions for rare decays that we
presented in \cite{BFRS-BIG}. In fact, as the central value of $q$ found in
the previous section is still on the high side from the
point of view of rare decays (compare (\ref{q-det}) with (\ref{qwithRD})), 
we expect that $R_{\rm c}$ and $R_{\rm n}$
will still slightly move towards each other, when the data
improve in the future. However, the most interesting question is whether 
the large negative values of $\phi$ and $\theta_i$ will be reinforced 
by the future more accurate data. This would be a very spectacular
signal of NP!

\section{Outlook}
We have summarized our strategy for analyzing $B\to \pi \pi $, $B\to \pi K$ 
decays and rare $K$ and $B$ decays. Within a simple NP scenario of enhanced  
CP-violating EW penguins considered by us, the NP contributions enter 
significantly only in certain $B\to\pi K$ decays and rare $K$ and $B$ decays, 
while the $B\to\pi\pi$ system is practically 
unaffected by these contributions and can be described within the SM. 
The confrontation of our strategy with the most recent data on 
$B \to \pi \pi$ and $B \to \pi K$ modes from the BaBar and Belle 
collaborations is very encouraging. 
In particular, our earlier predictions for the direct CP asymmetries of 
$B_d \to \pi^0 \pi^0$ and $B_d \to \pi^{\mp} K^{\pm}$ have been confirmed 
within the theoretical and experimental uncertainties, and the shift in the
experimental values of $R_{\rm c}$ and $R_{\rm n}$ took place as
expected. 

It will be exciting to follow the experimental progress on
$B\to\pi\pi$ and $B\to\pi K$ decays and the corresponding efforts in rare
decays. In particular, new messages from BaBar and Belle that 
the present central values of $R_{\rm c}$ and $R_{\rm n}$ have been confirmed 
at a high confidence level, a slight increase of $R$ and a message 
from KEK {\cite{KEK}} in the next two
years that the decay $K_{\rm L}\to\pi^0\nu\bar\nu$ has been observed would 
give a strong support to the NP scenario considered here.

\vspace*{0.5truecm}

\noindent
{\bf Acknowledgments}\\
\noindent
The work presented here was supported in part by the German 
Bundesministerium f\"ur
Bildung und Forschung under the contract 05HT4WOA/3.


\begin{thebibliography}{99}
%
%
%


\bibitem{BFRS-PRL}A.J. Buras, R. Fleischer, S. Recksiegel and F. Schwab,
%``B $\to$ pi pi, new physics in B $\to$ pi K and implications for rare K and B
%decays,''
{\it Phys.\ Rev.\ Lett.}~{\bf 92} (2004) 101804.
%[arXiv:hep-ph/0312259].
%%CITATION = HEP-PH 0312259;%%

\bibitem{BFRS-BIG}A.J. Buras, R. Fleischer, S. Recksiegel and F. Schwab,
%``Anatomy of prominent B and K decays and signatures of CP-violating new
%physics in the electroweak penguin sector,''
{\it Nucl.\ Phys.}~{\bf B697} (2004) 133.
%arXiv:hep-ph/0402112.
%%CITATION = HEP-PH 0402112;%%

\bibitem{BFRS-UPDATE}
A.~J.~Buras, R.~Fleischer, S.~Recksiegel and F.~Schwab,
%``The B $\to$ pi pi, pi K puzzles in the light of new data: Implications for
%the standard model, new physics and rare decays,''
hep-ph/0410407.
%%CITATION = HEP-PH 0410407;%%

\bibitem{Buras:1998ed}
A.J. Buras and L. Silvestrini,
%``Upper bounds on K $\to$ pi nu anti-nu and K(L) $\to$ pi0 e+ e- from  
%epsilon'/epsilon and K(L) $\to$ mu+ mu-,''
{\it Nucl.\ Phys.}~{\bf B546} (1999) 299.
%[arXiv:hep-ph/9811471].
%%CITATION = HEP-PH 9811471;%%

\bibitem{BRS}A.J. Buras, A. Romanino and L. Silvestrini, 
{\it Nucl.\ Phys.}~{\bf B520} (1998) 3.

\bibitem{Buras:1999da}
A.J. Buras, G. Colangelo, G. Isidori, A. Romanino and L. Silvestrini,
%``Connections between epsilon'/epsilon and rare kaon decays in  
%supersymmetry,''
{\it Nucl.\ Phys.}~{\bf B566} (2000) 3.
%[arXiv:hep-ph/9908371].
%%CITATION = HEP-PH 9908371;%%

\bibitem{Buchalla:2000sk}
G. Buchalla, G. Hiller and G. Isidori,
%``Phenomenology of non-standard Z couplings in exclusive semileptonic  
%b $\to$ s transitions,''
{\it Phys.\ Rev.}~{\bf D63} (2001) 014015;\\
%[arXiv:hep-ph/0006136];
%%CITATION = HEP-PH 0006136;%%
D. Atwood and G. Hiller,
%``Implications of non-standard CP violation in hadronic B decays,''
LMU-09-03 [hep-ph/0307251].
%%CITATION = HEP-PH 0307251;%%


\bibitem{BaBar-Bpi0pi0}B. Aubert {\it et al.}\  [BaBar Collaboration],
%``Study of B0 (anti-B0) $\to$ pi0 pi0, B+- $\to$ pi+- pi0 and 
%B+- $\to$ K+- pi0 decays,''
BABAR-CONF-04/035 [hep-ex/0408081].
%%CITATION = HEP-EX 0408081;%%

\bibitem{Chao:2003ue}
Y.~Chao {\it et al.}  [Belle Collaboration],
%``Improved measurements of branching fractions for B $\to$ K pi, pi pi and K
%anti-K decays,''
{\it Phys.\ Rev.}~{\bf D69} (2004) 111102.
%[arXiv:hep-ex/0311061].
%%CITATION = HEP-EX 0311061;%%


\bibitem{Aubert:2002jb}
B.~Aubert {\it et al.}  [BABAR Collaboration],
%``Measurements of branching fractions and CP-violating asymmetries in B0 $\to$
%pi+ pi-, K+ pi-, K+ K- decays. ((B)),''
{\it Phys.\ Rev.\ Lett.}~{\bf 89} (2002) 281802.
%[arXiv:hep-ex/0207055].
%%CITATION = HEP-EX 0207055;%%


\bibitem{Belle-Bpi0pi0-new}K.~Abe {\it et al.}\  [Belle Collaboration],
%``Observation of B0 $\to$ pi0 pi0,''
BELLE-CONF-0406 [hep-ex/0408101].
%%CITATION = HEP-EX 0408101;%%

\bibitem{BaBar-CP-Bpipi}B. Aubert {\it et al.}\ [BaBar Collaboration],
BABAR-CONF-04/047 [hep-ex/0408089].
%%CITATION = HEP-EX 0408089;%%

\bibitem{Belle-CP-Bpipi}K. Abe {\it et al.}\ [Belle Collaboration],
{\it Phys.\ Rev.\ Lett.}~{\bf 93} (2004) 021601.
%%CITATION = HEP-EX 0401029;%%

\bibitem{HFAG}Heavy Flavour Averaging Group,
{\tt http://www.slac.stanford.edu/xorg/hfag/}.

\bibitem{PDG}S. Eidelman {\it et al.}\ [Particle Data Group], 
{\it Phys.\ Lett.}~{\bf B592} (2004) 1.

%%%

\bibitem{BBNS}
M.~Beneke, G.~Buchalla, M.~Neubert and C.~T.~Sachrajda,
%``{QCD} factorization for B $\to$ pi pi decays: Strong phases and CP  violation
%in the heavy quark limit,''
{\it Phys.\ Rev.\ Lett.\ }~{\bf 83} (1999) 1914;\\
%[arXiv:hep-ph/9905312].
%%CITATION = HEP-PH 9905312;%%
M.~Beneke, G.~Buchalla, M.~Neubert and C.~T.~Sachrajda,
%``QCD factorization for exclusive, non-leptonic B meson decays: General
%arguments and the case of heavy-light final states,''
{\it Nucl.\ Phys.\ B }~{\bf 591} (2000) 313.
%[arXiv:hep-ph/0006124].
%%CITATION = HEP-PH 0006124;%%

\bibitem{Be-Ne}M. Beneke and M. Neubert,
%``QCD factorisation for B $\to$ P P and B $\to$ P V decays,''
{\it Nucl.\ Phys.}~{\bf B675} (2003) 333.
%[arXiv:hep-ph/0308039].
%%CITATION = HEP-PH 0308039;%%

\bibitem{WO}
L. Wolfenstein, {\it Phys.\ Rev.\ Lett.}~{\bf 51} (1983) 1945.

\bibitem{BLO}
{ A.J. Buras, M.E. Lautenbacher and G. Ostermaier,}
{\it Phys.\ Rev.}~{\bf D50} (1994) 3433.

\bibitem{PAP0}A.J. Buras and R. Fleischer,
%``Limitations in measuring the angle Beta by using SU(3) relations for 
%B meson decay amplitudes,''
{\it Phys.\ Lett.}~{\bf B341} (1995) 379.
%[arXiv:hep-ph/9409244].
%%CITATION = HEP-PH 9409244;%%

\bibitem{c-pen}
M. Ciuchini, E. Franco, G. Martinelli, L. Silvestrini,
%``Charming penguins in B decays,''
{\it Nucl.\ Phys.}~{\bf B501} (1997) 271;\\
%[arXiv:hep-ph/9703353].
%%CITATION = HEP-PH 9703353;%%
C. Isola, M. Ladisa, G. Nardulli, T.N. Pham and P. Santorelli,
%``Charming penguin contributions to B $\to$ K pi,''
%%CITATION = HEP-PH 0101118;%%
%``Charming penguin contributions to charmless B decays into two  
%pseudoscalar mesons,''
{\it Phys.\ Rev.}~{\bf D64} (2001) 014029 and {\bf D65} (2002) 094005;\\
%%CITATION = HEP-PH 0110411;%%
M. Ciuchini, E. Franco, G. Martinelli, M. Pierini and 
L. Silvestrini,
%``Charming penguins strike back,''
{\it Phys.\ Lett.}~{\bf B515} (2001) 33.
%[arXiv:hep-ph/0104126].
%%CITATION = HEP-PH 0104126;%%

\bibitem{BFM}A.J. Buras, R. Fleischer and T. Mannel,
%``Penguin topologies, rescattering effects and penguin hunting with  B(u,d)
%$\to$ K anti-K and B+- $\to$ pi+- K,''
{\it Nucl.\ Phys.}~{\bf B533} (1998) 3.
%[arXiv:hep-ph/9711262].
%%CITATION = HEP-PH 9711262;%%

\bibitem{BPRS}C.W. Bauer, D. Pirjol, I.Z. Rothstein and I.W. Stewart,
%``B $\to$ M(1) M(2): Factorization, charming penguins, strong phases, and
%polarization,''
hep-ph/0401188.
%%CITATION = HEP-PH 0401188;%%

\bibitem{FR-I}R. Fleischer and S. Recksiegel,
%``Waiting for the discovery of B/d0 $\to$ K0 anti-K0,''
{\it Eur.\ Phys.\ J.}~{\bf C} (2004), Online First, 
DOI: 10.1140/epjc/s2004-02023-0 (hep-ph/0408016).
%%CITATION = HEP-PH 0408016;%%

\bibitem{ALP-Bpipi}A. Ali, E. Lunghi and A.Y. Parkhomenko,
%``An analysis of the time-dependent CP asymmetry in B $\to$ pi pi 
%decays in the
%standard model,''
{\it Eur.\ Phys.\ J.}~{\bf C36} (2004) 183.
%[arXiv:hep-ph/0403275].
%%CITATION = HEP-PH 0403275;%%

\bibitem{CGRS}C.W. Chiang, M. Gronau, J.L. Rosner and D.A. Suprun,
%``Charmless B $\to$ P P decays using flavor SU(3) symmetry,''
{\it Phys.\ Rev.}~{\bf D70} (2004) 034020.
%[arXiv:hep-ph/0404073].
%%CITATION = HEP-PH 0404073;%%

\bibitem{He:2004ck}
X.~G.~He and B.~H.~J.~McKellar,
%``Hadron decay amplitudes from B $\to$ K pi and B $\to$ pi pi decays,''
hep-ph/0410098.
%%CITATION = HEP-PH 0410098;%%

%\cite{Bauer:2004ck}
\bibitem{Bauer:2004ck}
C.~W.~Bauer and D.~Pirjol,
%``Graphical amplitudes from SCET,''
{\it Phys.\ Lett.\ B}~{\bf 604} (2004) 183.
%[arXiv:hep-ph/0408161].
%%CITATION = HEP-PH 0408161;%%

%\cite{Feldmann:2004mg}
\bibitem{Feldmann:2004mg}
T.~Feldmann and T.~Hurth,
%``Non-factorizable contributions to B $\to$ pi pi decays,''
hep-ph/0408188.
%%CITATION = HEP-PH 0408188;%%

\bibitem{BaBar-BdKK-obs}B. Aubert {\it et al.}\ [BaBar Collaboration],
%``Measurements of branching fractions and CP-violating asymmetries in B-meson
%decays to the charmless two-body states K0 pi+, anti-K0 K+, and K0 anti-K0,''
BABAR-CONF-04/044 [hep-ex/0408080].
%%CITATION = HEP-EX 0408080;%%

\bibitem{BaBar-BK0pi0}B. Aubert {\it et al.}\ [BaBar Collaboration],
%``Measurements of the branching fraction and CP-violating asymmetries of B0
%$\to$ K0(S) pi0 decays,''
BABAR-CONF-04/30 [hep-ex/0408062].
%%CITATION = HEP-EX 0408062;%%

\bibitem{BF-neutral2}A.J. Buras and R. Fleischer,
%``Constraints on the CKM angle gamma and strong phases from B $\to$ pi K  
%decays,''
{\it Eur.\ Phys.\ J.}~{\bf C16} (2000) 97.
%%CITATION = HEP-PH 0003323;%%

\bibitem{BaBar-CP-dir-obs}B. Aubert {\it et al.}\  [BaBar Collaboration],
%``Observation of direct CP violation in B0 $\to$ K+ pi- decays,''
{\it Phys.\ Rev.\ Lett.}~{\bf 93} (2004) 131801.
%hep-ex/0407057.
%%CITATION = HEP-EX 0407057;%%

\bibitem{Belle-CP-dir-obs}Y. Chao {\it et al.}\  [Belle Collaboration],
%``Evidence for direct CP violation in B0 $\to$ K+ pi- decays,''
{\it Phys.\ Rev.\ Lett.}~{\bf 93} (2004) 191802.
%hep-ex/0408100.
%%CITATION = HEP-EX 0408100;%%

\bibitem{RF-Bpipi}R. Fleischer,
%``Constraining penguin contributions and the CKM angle gamma through  B/d
%$\to$ pi+ pi-,''
{\it Eur.\ Phys.\ J.}~{\bf C16} (2000) 87.
%[arXiv:hep-ph/0001253].
%%CITATION = HEP-PH 0001253;%%

\bibitem{Fl-Ma}R. Fleischer and J. Matias,
%``Exploring CP violation through correlations in B $\to$ pi K,  B/d $\to$ pi+
%pi-, B/s $\to$ K+ K- observable space,''
{\it Phys.\ Rev.}~{\bf D66} (2002) 054009.
%[arXiv:hep-ph/0204101].
%%CITATION = HEP-PH 0204101;%%

\bibitem{BSU}A.J. Buras, F. Schwab and S. Uhlig,
%``Waiting for precise measurements of K+ $\to$ pi+ nu anti-nu and K(L) $\to$
%pi0 nu anti-nu,''
TUM-HEP-547 [hep-ph/0405132].
%%CITATION = HEP-PH 0405132;%%

\bibitem{Wu:2004xx}
Y.~L.~Wu and Y.~F.~Zhou,
%``Implications of charmless B decays with large direct CP violation,''
hep-ph/0409221.
%%CITATION = HEP-PH 0409221;%%

\bibitem{FM}R. Fleischer and T. Mannel,
%``Constraining the CKM angle gamma and penguin contributions through  combined
%B $\to$ pi K branching ratios,''
{\it Phys.\ Rev.}~{\bf D57} (1998) 2752.
%[arXiv:hep-ph/9704423].
%%CITATION = HEP-PH 9704423;%%

\bibitem{NR}M. Neubert and J.L. Rosner,
%``New bound on gamma from B+- $\to$ pi K decays,''
{\it Phys.\ Lett.}~{\bf B441} (1998) 403;
%%CITATION = HEP-PH 9808493;%%
%``Determination of the weak phase gamma from rate measurements in  
%B+- $\to$ pi K, pi pi decays,''
{\it Phys.\ Rev.\ Lett.}~{\bf 81} (1998) 5076.
%%CITATION = HEP-PH 9809311;%%

\bibitem{Barger:2004hn}
V.~Barger, C.~W.~Chiang, P.~Langacker and H.~S.~Lee,
%``Solution to the B $\to$ pi K puzzle in a flavor-changing Z' model,''
{\it Phys.\ Lett.}{\bf B598}, (2004) 218
%[arXiv:hep-ph/0406126].
%%CITATION = HEP-PH 0406126;%%

\bibitem{BFRS-I}A.J. Buras, R. Fleischer, S. Recksiegel and F. Schwab,
%``The B $\to$ pi K puzzle and its relation to rare B and K decays,''
{\it Eur.\ Phys.\ J.}~{\bf C32} (2003) 45.
%[arXiv:hep-ph/0309012].
%%CITATION = HEP-PH 0309012;%%

\bibitem{Kaneko:2002mr}
J.~Kaneko {\it et al.}  [Belle Collaboration],
%``Measurement of the electroweak penguin process B $\to$ X/s l+ l-. ((B)),''
{\it Phys.\ Rev.\ Lett. }  {\bf 90} (2003) 021801
%[arXiv:hep-ex/0208029].
%%CITATION = HEP-EX 0208029;%%

\bibitem{Abe:2004sg}
K.~Abe {\it et al.}  [Belle Collaboration],
%``Improved measurement of the electroweak penguin process B $\to$ X/s l+ l-,''
hep-ex/0408119.
%%CITATION = HEP-EX 0408119;%%

\bibitem{Buras:2000dm}
A.~J.~Buras, P.~Gambino, M.~Gorbahn, S.~J{\"a}ger and L.~Silvestrini,
%``Universal unitarity triangle and physics beyond the standard model,''
{\it Phys.\ Lett.}{\bf B500} (2001) 161
%[arXiv:hep-ph/0007085].
%%CITATION = HEP-PH 0007085;%%

\bibitem{D'Ambrosio:2002ex}
G.~D'Ambrosio, G.~F.~Giudice, G.~Isidori and A.~Strumia,
%``Minimal flavour violation: An effective field theory approach,''
{\it Nucl.\ Phys.}  {\bf B645} (2002) 155
%[arXiv:hep-ph/0207036].
%%CITATION = HEP-PH 0207036;%%


\bibitem{BF-MFV}A.J. Buras and R. Fleischer,
%``Bounds on the unitarity triangle, sin(2beta) and K $\to$ pi nu anti-nu
%decays in models with minimal flavor violation,''
{\it Phys.\ Rev.}~{\bf D64} (2001) 115010.
%[arXiv:hep-ph/0104238].
%%CITATION = HEP-PH 0104238;%%

\bibitem{BBSIN}
{ G. Buchalla and A.J. Buras}, 
{\it Phys.\ Lett.}~{\bf B333} (1994) 221,
{\it Phys.\ Rev.}~{\bf D54} (1996) 6782.

\bibitem{Buras:2004qb}
A.~J.~Buras, T.~Ewerth, S.~J\"ager and J.~Rosiek,
%``K+ $\to$ pi+ nu anti-nu and K(L) $\to$ pi0 nu anti-nu decays in the general
%MSSM,''
hep-ph/0408142.
%%CITATION = HEP-PH 0408142;%%

\bibitem{GRNR}
%\cite{Grossman:1997sk}
%\bibitem{Grossman:1997sk}
Y. Grossman and Y. Nir,
%``K(L) $\to$ pi0 nu anti-nu beyond the standard model,''
{\it Phys.\ Lett.}~{\bf B398} (1997) 163.
%[arXiv:hep-ph/9701313].
%%CITATION = HEP-PH 9701313;%%

%\cite{Anisimovsky:2004hr}
\bibitem{Anisimovsky:2004hr}
V.~V.~Anisimovsky {\it et al.}  [E949 Collaboration],
%``Further study of the decay K+ $\to$ pi+ nu anti-nu,''
{\it Phys.\ Rev.\ Lett.\ }{\bf 93} (2004) 031801.
%[arXiv:hep-ex/0403036].
%%CITATION = HEP-EX 0403036;%%

\bibitem{Alavi-Harati:1999hd}
A.~Alavi-Harati {\it et al.}  [The E799-II/KTeV Collaboration],
%``Search for the decay K(L) $\to$ pi0 nu anti-nu using pi0 $\to$ e+ e-
%gamma,''
{\it Phys.\ Rev.\ }{\bf D61} (2000) 072006.
%[arXiv:hep-ex/9907014].
%%CITATION = HEP-EX 9907014;%%

\bibitem{Alavi-Harati:2003mr}
A.~Alavi-Harati {\it et al.}  [KTeV Collaboration],
%``Search for the rare decay K(L) $\to$ pi0 e+ e-,''
{\it Phys.\ Rev.\ Lett.\ }{\bf 93} (2004) 021805.
%[arXiv:hep-ex/0309072].
%%CITATION = HEP-EX 0309072;%%

\bibitem{Alavi-Harati:2000hs}
A.~Alavi-Harati {\it et al.}  [KTeV Collaboration],
%``Search for the decay K(L) $\to$ pi0 mu+ mu-,''
{\it Phys.\ Rev.\ Lett.\ }{\bf 84} (2000) 5279.
%[arXiv:hep-ex/0001006].
%%CITATION = HEP-EX 0001006;%%

\bibitem{Barate:2000rc}
R.~Barate {\it et al.}  [ALEPH Collaboration],
% ``Measurements of BR(b $\to$ tau- anti-nu/tau X) and BR(b $\to$ tau-
%anti-nu/tau D*+- X) and upper limits on BR(B- $\to$ tau- anti-nu/tau)  
%and BR(b%$\to$ s nu anti-nu),''
{\it Eur.\ Phys.\ J.\ }{\bf C19} (2001) 213.
%[arXiv:hep-ex/0010022].
%%CITATION = HEP-EX 0010022;%%

\bibitem{Abazov:2004dj}
V.~M.~Abazov  [D0 Collaboration],
%``A Search for the Flavor-Changing Neutral Current Decay B0_s $\to$mu^+mu^-
%in pp(bar) Collisions at \sqrt{s} = 1.96 TeV with the D0 Detector,''
FERMILAB-PUB-04-215-E [hep-ex/0410039].
%%CITATION = HEP-EX 0410039;%%

\bibitem{Isidori:2004rb}
G.~Isidori, C.~Smith and R.~Unterdorfer,
%``The rare decay K(L) $\to$ pi0 mu+ mu- within the SM,''
{\it Eur.\ Phys.\ J.\ }{\bf C36} (2004) 57.
%[arXiv:hep-ph/0404127].
%%CITATION = HEP-PH 0404127;%%

\bibitem{KEK}
J-PARC, http://www-ps.kek.jp/jhf-np/LOIlist/LOIlist.html

\end{thebibliography}
\end{document}